\documentclass[aps,prb,twocolumn,showpacs]{revtex4}

\usepackage{amsmath}
\usepackage{color}
\usepackage{amssymb}

\usepackage{graphicx}

\begin{document}

\title{URu$_2$Si$_2$: hidden order and  amplitude of quantum oscillations}

\author{V.P.Mineev}
\affiliation{Commissariat a l'Energie Atomique, INAC / SPSMS, 38054 Grenoble, France}

\begin{abstract}
It is shown that the hidden ordered (HO)  state in  heavy fermionic metal URu$_2$Si$_2$  is either nonconventional charge density wave (CDW)   or antiferroelectric (AFE)  commensurate ordering.  Similar to antiferromagnetic  (AF) order an  antiferroelectric order creates  momentum dependence of conducting electrons $g$-factor. This even modest anisotropy along with the anisotropy of heavy cyclotron mass  produce multiple effect of so called zero-spin splitting  in the amplitude of the de Haas - van Alphen and the Shubnikov - de Haas signals as function of direction of magnetic field.
\end{abstract}

\date{\today}
\maketitle

\section{Introduction}

During the past three decades, the puzzle of hidden order (HO)  in URu$_2$Si$_2$ attracts attention of condensed matter community (for the recent reviews see \cite{Mydosh2014, Mydosh2011} and references therein). This strange state develops below $17.5 K$. The corresponding electronic specific heat measurements indicates that below this  second-order-type transition about one half of the Fermi surface is removed. Nevertheless the material is still a good metal passing to the superconducting state below 1.4 K.

The de Haas van Alphen oscillations corresponding to three kinds of Fermi surfaces were observed clearly in both the normal and superconducting mixed states of URu$_2$Si$_2$ \cite{Ohkuni1999}. The frequency of oscillations 
of so called $\alpha$ band is almost  independent from the field direction that corresponds to the practically spherical Fermi surface. At the same time the dHvA amplitude does not change with changes of field direction in the basal tetragonal plane while in the (010) plane, the signal amplitude becomes zero at about 10 different field angles.
 The authors of Ref.10  relate this observation with phenomenon of so called zero spin-splitting in which the up and down spin contribution to the oscillation cancel out. Namely,
 the dHvA amplitude is proportional to the spin factor 
 \begin{equation}
 \cos(\pi g m^\star_c/2m)
 \label{cos}
 \end{equation}
  depending of $g$-factor value and cyclotron effective mass $m^\star_c$.\cite{Shoenberg,
 Pit} 
The latter measured from the Dingle plot is 
 about 10 times larger than the bar electron mass $m$. It means that  $g$-factor  undergoes   noticable variation as function of the field direction changed from parallel to tetragonal axis to  parallel to a-axis. 

We note, however, that the  effect of multiple nullification of the dHvA signal amplitude as  function of magnetic field direction is determined by the product of g-factor and the cyclotron mass. These are quantities averaged over the line limiting the extremal cross section of the Fermi surface. So, in heavy fermionic materials when $m^*>>m$ , even not so strong anisotropy in each of these
 quantities  can produce multiple nullification in the dHvA amplitude as function of direction of magnetic field.

 This phenomenon quite recently has been studied  by measurements of Shubnikov-de Haas oscillations in the same material.\cite{Aoki} In addition to previous observations \cite{Ohkuni1999} there was revealed the  similar angular dependence of amplitude of $\beta$  band. The measured \cite{Hassinger2010} cyclotron masses
 for magnetic field parallel to $c$-axis are: $m_\alpha^*=12.4 m$ and $m_\beta^*=23.8 m$. They are proved roughly twice smaller for field parallel to $a$ axis. So, if the g-factor magnitude  undergoes similar double or so decrease,  the multiple nullification of dHvA signal amplitude is quite understandable.
 
 The g-factor anisotropy arises in a multiband metal possessing centre symmetry due to interband spin-orbital coupling. In case of tetragonal URu$_2$Si$_2$ the g-factor presents an uniaxial tensor
 \begin{equation}
 g_{\alpha\beta}({\bf k})=g_\perp({\bf k})(\hat x_\alpha\hat x_\beta+\hat y_\alpha\hat y_\beta)
+g_\parallel({\bf k})\hat z_\alpha\hat z_\beta.
 \end{equation}
 So, the  modest scale  of  g-factor anisotropy can be expectable due to spin-orbital coupling without attracting of exotic mechanisms  in an attempt to explain seeming Ising anisotropy of g-factors of  conducting electrons.

 Effective $g$-factor value  determining amplitude of de Haas-van Alphen signal  from the particular extremal cross section of the Fermi surface is given by the average the momentum dependent $g$-factor over the curve encircling this area
 \begin{equation}
 g_{eff}=\frac{\oint \frac{dl}{v_F({\bf k})}g({\bf k})}{\oint \frac{dl}{v_F({\bf k})}}.
 \end{equation}
 The  anisotropy of $g_{eff}$  has no direct relationship with the $g$-factor anisotropy determining so called paramagnetic limiting field for superconducting state as it was proposed in recent publications \cite{Alt}. 
The latter is determined by  $g$-factors  extracted from the  magnetic susceptibility proportional to the average of the product of square of local $g({\bf k})$ factor and the local density of states  $N_0({\bf k})$ over the whole  Fermi surface area including all band sheets.  The whole susceptibility can also include the van Vleck component.
So, the  anisotropy in "$g$-factor"  extracted from susceptibility has less common with anisotropy of de Haas-van Alphen $g$-factor. 
It is also pertinent to mention here, that  although the susceptibility in URu$_2$Si$_2$ along $c$-axis is three times larger than along $a$-axis \cite{Palstra}
it is not an argument to think  that the suppression of superconducting state in URu$_2$Si$_2$ is determined by pure paramagnetic mechanism.

An attempt to explain the $g$-factor anisotropy  has been undertaken by virtue of arising  of "hastatic" ordering that breaks double time-reversal symmetry, mixing states of integer and half-spin.\cite{Flint} 
 
  In contrast to this exotic approach recently on the basis of relativistic density functional theory  there was shown  that the bandlike $5f$ electrons in URu$_2$Si$_2$ exhibit colossal Ising behavior.\cite{Werwinski} The origin of the peculiar anisotropy is found due to specific Fermi surface structure and the strong spin-orbital interaction. 
The calculations \cite{Werwinski} has been performed for the AF phase where  according to Ref.16 the Fermi surface is practically identical to that of the HO phase. 

We have already argued
that multiple nullification of dHvA amplitude in a 
 heavy fermionic metal can be explained  taking  into account  some modest anisotropy of effective $g$-factor arising due to interband spin-orbital interaction.
Here we will show that the g-factor anisotropy  inevitably appears even in a single band metal  due to formation of specific commensurate charge density wave. In the next section  it is shown that  is  the most plausible candidate for the URu$_2$Si$_2$ hidden ordered state  is commensurate antiferroelectric ordering.
The source of $g$-factor anisotropy based on argumentation  has been proposed several years ago by 
 R.Ramazashvili \cite{Ramazashvili2008,Ramazashvili2009}  that a commensurate  antiferromagnetic order  leads to  momentum dependence of conducting  electrons $g$-factor.
 One can come to  similar conclusions for a metal with antiferroelectric ordering as well. This will be done in the third section followed by conclusion.
 
\section{Hidden Order}

There was proposed a lot of theories to explain the HO phenomenon (see reviews  \cite{Mydosh2014, Mydosh2011}). Meanwhile  several recent and not so recent experimental developments are able to put serious restrictions on theoretical phantasies. In fact  the symmetry  of HO state is practically fixed   by experimental observations. 

URu$_2$Si$_2$ is a tetragonal material with uranium atoms forming body centered tetragonal lattice. There was found  \cite{Amitsuka2009} that  at pressure 0.5 GPa 
at low enough  temperatures  HO state  abruptly  transforms to two-sub-lattice antiferromagnetic state.  The magnetic moments of the order 0.4 $\mu_B$
aligned parallel and antiparallel to $c$-axis are  disposed on uranium atoms with ordering vector ${\bf q}=(1,0,0)$ such that the body-centered tetragonal structure transforms to the simple tetragonal one. Further investigations demonstrated that  at all temperatures HO and AF state are separated from each other  by the first order type transition,  that is the first order transition line is finished at  some bicritical point on the line of the second-order-type transition $T_c$$($$P$$)$.\cite{Motoyama2003} This observation has the important consequence:  the HO state and AF state have to  have the different symmetries. This is because at  their symmetry coincidence,  the line of the first order transition should obligatory terminate at some critical point  below $T_c$$($$P$$)$ as it was demonstrated in the paper \cite{Mineev2005}.

Another important observation relating to symmetry of HO state  was done by measurements of Shubnikov-de-Haas effect on high quality URu$_2$Si$_2$  single crystals.\cite{Hassinger2010} Namely, there was shown that under pressure for field $H\parallel \hat z$ the Fermi surface reveals only minor changes between the HO state and AF state.
This was strong indication that both phases have the same unit cell doubling and the same ordering vector. 

The change in the electronic periodicity at the transition from the normal body-centered tetragonal state to the simple tetragonal HO state   has been revealed  recently by ARPES \cite{Meng2013,Yoshida2013,Bareille2014} and polarization resolved Raman spectroscopy \cite{Kung2014,Buhot2014} measurements.  
Thus, the similarity  between HO phase and the high pressure AF phase found in quantum-oscillation experiments has been confirmed.

So, we come to the conclusion  that   normal and HO states have different translational symmetries and the HO and AF states have the same translational and rotational symmetry but at the same time these phases should be symmetrically different. This case, there  is only one opportunity for hidden ordering. It   is nonconventional commensurate charge density wave (CDW) along vector ${\bf q}=(1,0,0)$, that is periodic in space ordering of multipole charge distributions  around the uranium sites possessing of local tetragonal symmetry.  The appearance of such type ordering has to  change magnetization distribution induced in a single crystal of URu$_2$Si$_2$ under a magnetic field applied along the tetragonal c axis that has been observed and  reported in Ref.22.

The nonconventional commensurate spin density  wave (SDW) as candidate for the HO is forbidden because this case the HO and AF phases will have the same symmetry. 
On the contrary the symmetry allows nonconventional CDW ordering including  local breaking of the space parity  like a periodic distribution of multipoles alternating by their  mirror reflections.  This case the initial normal state body centered  tetragonal lattice transforms below 17.5 K to simple tetragonal material formed by two sub-lattices differing from  each other by the space inversion.
Such type commensurate chiral density wave ground state has actually been proposed in Ref. 20. We shall call this state  antiferroelectric state (AFE).  This state can  in general  include also the usual antiferromagnetic component such that two sub-lattices transfer each other by application both space and time inversion and shift on vector  ${\bf q}=(1,0,0)$.

Thus,  the HO state is either nonconventional CDW  or AFE commensurate ordering. In the former case we deal with  the simple doubling of body-centered tetragonal unit cell modulated by charge distribution, 
in the  latter,  the body-centered  lattice consists of two
 sub-lattices differ each other at least by the space parity transformation. We shall demonstrate that  this gives rise to
 momentum dependence of electron $g$-factor revealing itself in the  dependence of the amplitude of quantum magnetic oscillations from the direction of magnetic field.

 \section{g-factor anisotropy in antiferroelectric state}
 
 The modification of electron spectrum in single band metal caused  by an antiferroelectric (AFE) ordering doubling the initial period of crystal lattice can be derived introducing the Rashba-Bychkov modulation in one-electron Hamiltonian
 \begin{eqnarray}
H_{AFE}=\sum_{{\bf k}}\left [(\varepsilon_{\bf k}\delta_{\alpha\beta}
-{\bf h}\mbox{\boldmath$\sigma$}_{\alpha\beta})
a^\dagger_{{\bf k}\alpha}a_{{\bf k}\beta}\right.~~~~~~~~~~\nonumber\\ 
\left.+
i{\bf l}_{\bf k}
\mbox{\boldmath$\sigma$}_{\alpha\beta} (a^\dagger_{{\bf k}+{\bf q}/2,\alpha}a_{{\bf k}-{\bf q}/2,\beta}-a^\dagger_{{\bf k}-{\bf q}/2,\alpha}a_{{\bf k}+{\bf q}/2,\beta})\right ]
.
\label{1}
\end{eqnarray}
Here, 
\begin{equation}
{\bf h}=g\mu_B{\bf H}/2 
\end{equation}
 is constant magnetic field acting on electron spins, $ \mbox{\boldmath$\sigma$}  $ are the Pauli matrices. For simplicity we keep only one harmonic in AFE modulation. Its amplitude is determined by  real pseudovector ${\bf l}_{\bf k}$ which satisfies ${\bf l}_{\bf k}=-{\bf l}_{-{\bf k}}$ and $g{\bf l}_{g^{-1}{\bf k}}={\bf l}_{\bf k} $  where $g$  is any symmetry operation in the point group  $D_4$ not ncluding space inversion.

It is instructive to compare the hamiltonian (\ref{1}) with corresponding hamiltonian of  a single-band metal with single harmonic antiferromagnetic (AF) modulation  
 \begin{eqnarray}
H_{AF}=\sum_{{\bf k}}\left [(\varepsilon_{\bf k}\delta_{\alpha\beta}
-{\bf h}\mbox{\boldmath$\sigma$}_{\alpha\beta})
a^\dagger_{{\bf k}\alpha}a_{{\bf k}\beta}\right.~~~~~~~~~~\nonumber \\ 
\left.+
\mbox{\boldmath$\sigma$}_{\alpha\beta} 
({\bf h}^s_{\bf k}a^\dagger_{{\bf k}+{\bf q}/2,\alpha}a_{{\bf k}-{\bf q}/2,\beta}+{\bf h}^{s*}_{\bf k}a^\dagger_{{\bf k}-{\bf q}/2,\alpha}a_{{\bf k}+{\bf q}/2,\beta})\right ]
.
\label{2}
\end{eqnarray}
Here, the  amplitude of the staggered field ${\bf h}^s_{\bf k}={\bf h}^s_{-{\bf k}}$  is an even function of the momentum. 

The difference of two hamiltonians is the following : the AFE hamiltonian  is time reversal symmetric but it breaks the space inversion symmetry,  on the contrary, the AF hamiltonian breaks the time inversion but it keeps the space parity.\cite{Samokhin}
These difference, however,  is not important for electron energy bands dispersion: in both cases it is described by equivalent equations.

To find the band energies for AFE
 one must  diagonalize the energy matrix
\begin{equation}
\hat E=\left( \begin{array}{cc}
\varepsilon_{{\bf k}+{\bf q}/2}-{\bf h}\mbox{\boldmath$\sigma$} &i{\bf l}_{\bf k}\mbox{\boldmath$\sigma$}\\ 
-i{\bf l}_{\bf k}\mbox{\boldmath$\sigma$} &\varepsilon_{{\bf k}-{\bf q}/2}-{\bf h}\mbox{\boldmath$\sigma$} 
\end{array}\right ).
\end{equation}
 For simplicity we write the corresponding band energies for two particular field directions along and perpendicular tetragonal axis.
 
So, for ${\bf h} \parallel \hat z$ we obtain
\begin{equation}
{\cal E}_{h_z}=\varepsilon_+\pm\sqrt{{ l_x}^2+{ l_y}^2+ 
\left (\sqrt{\varepsilon_-^2+{\bf l}^2_z}\pm h_z\right )^2},
\label{z}
\end{equation}
and for ${\bf h}\parallel \hat x$ 
\begin{equation}
{\cal E}_{h_x}=\varepsilon_+\pm\sqrt{{ l_y}^2+{ l_z}^2+ 
\left (\sqrt{\varepsilon_-^2+{\bf l}^2_x}\pm h_x\right )^2}.
\label{p}
\end{equation}
Here
\begin{equation}
\varepsilon_\pm=\frac{\varepsilon_{{\bf k}+{\bf q}/2}\pm\varepsilon_{{\bf k}-{\bf q}/2}}{2}.
\end{equation}
Thus, the initial  band in field absence splits on two bands  due to AFE period doubling. Each of these bands splits on two under magnetic field. 

Let us assume that basal plane spin orbital coupling ${\bf l}^2_\perp= l_x^2+{ l_y}^2$ is negligibly small. Then for ${\bf h} \parallel \hat z$
 \begin{equation}
{\cal E}_{h_z}=\varepsilon_+\pm
\left (\sqrt{\varepsilon_-^2+{\bf l}^2_z}\pm h_z\right ),
\label{z1}
\end{equation}
and we see, that magnetic field band splitting does not undergo  a change, that means g-factor is completely the same as in absence of spin-orbital interaction:
\begin{equation}
g_\parallel=2.
\end{equation}
On the contrary for field ${\bf h}\parallel \hat x$ we have  in linear in field approximation 
\begin{equation}
{\cal E}_{h_x}\approx\varepsilon_+\pm
\left (\sqrt{\varepsilon_-^2+{\bf l}^2_z}\pm\frac{|\varepsilon_-|}{\sqrt{\varepsilon_-^2+{\bf l}^2_z}}h_x \right ).
\label{x1}
\end{equation}
If the magnitude $|\varepsilon_-|$ on  the corresponding   de Haas - van Alphen orbit  is much smaller than the spin-orbit  amplitude $|l_z|$  the spin splitting proves to be negligibbly small: 
\begin{equation}
g_\perp\ll g_\parallel.
\end{equation}
So, by comparison of equations (\ref{z1}) and (\ref{x1}) we come to the conclusion  of g-factor anisotropy.

\section{Conclusion}

The whole body of experimental developments including the de Haas - van Alphen and the Shubnikov- de Haas measurements, ARPES and Raman spectroscopy and polarized neutron scattering allows to  fixe the order parameter of so called hidden order in URu$_2$Si$_2$ as nonconventional charge density wave or antiferroelectric commensurate ordering.   It is shown that  like the antiferromagnetism the antiferroelectric modulation causes essential quasi-momentum dependence of gyromagnetic factor of conducting electrons.
Depending on particular band structure and spin-orbital interaction  the g-factor anisotropy can be strong or weak. However, in heavy fermionic metals even modest anisotropy  of g-factor  produce effect of zero-spin splitting  manifesting itself in multiple nullification in the amplitude of the de Haas - van Alphen and the Shubnikov - de Haas signals as function of direction of magnetic field.

\section*{Acknowledgements}

I express my gratitude to Dai Aoki who has kindly informed me about the recent Shubnikov - de Haas data.

\end{document}